\documentclass[fleqn,twoside]{article}
\usepackage{espcrc2}
\usepackage{amssymb}
\usepackage{latexsym}

\usepackage{graphicx}
\usepackage[figuresright]{rotating}

\def\bc{\begin{center}}
\def\ec{\end{center}}
\newcommand{\be}{\begin{equation}}
\newcommand{\ee}{\end{equation}}

\newcommand{\MSbar}{\overline{\mathrm{MS}}}

\newcommand{\beq}{\begin{equation}}
\newcommand{\eeq}{\end{equation}}
\newcommand{\beqs}{\begin{equation*}}
\newcommand{\eeqs}{\end{equation*}}
\newcommand{\beqn}{\begin{eqnarray}}
\newcommand{\eeqn}{\end{eqnarray}}
\newcommand{\bea}{\begin{eqnarray}}
\newcommand{\eea}{\end{eqnarray}}
\newcommand{\beqns}{\begin{eqnarray*}}
\newcommand{\eeqns}{\end{eqnarray*}}

\def\vdir{v\kern-5.75pt\raise0.15ex\hbox{${\scriptstyle /}$}}
\def\pdir{p\kern-7.8pt\raise0.2ex\hbox{\Big{/}}}
\def\ddir{D\kern-7pt\raise0.2ex\hbox{\big{/}}}
\def\partdir{\partial\kern-7.6pt\raise0.25ex\hbox{/}}
\def\ddirp{D_{\kern-2.75pt\perp}\kern-11pt\raise0.2ex\hbox{\big{/}}\kern+4.5pt}

\newcommand{\re}[1]{(\ref{#1})}

\newcommand{\AmS}{{\protect\the\textfont2
  A\kern-.1667em\lower.5ex\hbox{M}\kern-.125emS}}

\title{Non Perturbative Renormalization in Coordinate Space
\thanks{Presented by J.~Reyes at ``Lattice 2003", Tsukuba.}
\thanks{This work has been  supported in part by the EU IHP under HPRN-CT-2000-00145
Hadrons/LatticeQCD.}}

\author{
V.~Gim\'enez \address{Dep. de F\'\i sica Te\`orica and IFIC, Univ. de
	Val\`encia, Dr. Moliner 50, E-46100, Burjassot, Val\`encia, Spain.},
L.~Giusti  \address{Theory Division, CERN, CH-1211 Geneva 23, Switzerland}, 
S.~Guerriero \address[roma3]{Dip. di Fisica, Univ. di Roma Tre and INFN-Sezione di
	Roma III, Via della Vasca
	Navale 84, I-00146 Roma, Italy.},
V.~Lubicz\addressmark[roma3],
G.~Martinelli\address[roma1]{Dip. di Fisica, Univ. di Roma ``La Sapienza'', Piazzale A. Moro 2, I-00185 Roma, Italy.}
              \address[infn]{INFN-Sezione di Roma I, Piazzale A. Moro 2, I-00185 Roma, Italy.},
S.~Petrarca\addressmark[roma1]\ \addressmark[infn],
J.~Reyes\addressmark[roma1]\ \addressmark[infn],
B.~Taglienti\addressmark[infn],
E.~Trevigne \addressmark[roma1]}
\begin{document}

\begin{abstract}
\mbox{}\\[-0.5cm]
We present an exploratory study of a gauge-invariant non-perturbative
renormalization technique. The renormalization conditions are imposed on
correlation functions of composite operators in coordinate space on the lattice.
Numerical results for bilinears obtained with overlap and O(a)-improved Wilson
fermions are presented. 
The measurement of the quark condensate is also discussed.
\end{abstract}
\maketitle
Renormalization constants (RC) of lattice operators are crucial inputs to obtain
physical quantities from the matrix elements of these operators. 
We have presented in this conference a numerical study of a 
non-perturbative (NP) technique proposed in \cite{Martinelli:1997zc}. This method has
the advantages of being gauge-invariant and free of contact terms, in
addition with the very easy implementation on a simulation. The first two
properties can be of vital interest to renormalize the operators which enter the
$K\to\pi\,\pi$ decays.

A very
exploratory analysis was already presented in the Lattice 2002 conference
\cite{Becirevic:2002yv} for bilinears operators. The results presented 
there were not conclusive 
and claimed for a more detailed study, which is the purpose of this
contribution. We have considered only bilinears operators to test the method,
since they are the simplest operators for which there exist results from
others NP techniques to compare
with.

The method consists on imposing renormalization conditions on the bare green
functions, $G_\Gamma(x)\equiv \langle O_\Gamma(x) O_\Gamma^\dagger(0) \rangle$,
where $O_\Gamma(x)$ is a quark bilinear operator, 
$O_\Gamma(x)=\bar \psi_1(x)\, \mathrm{\Gamma} \psi_2(x)$. One can choose
\be
Z^2_\Gamma(\mu=1/x)\, G_\Gamma(x)\,=\, 
 G_\Gamma(x)^{(free\: cont.)}\:,
\ee
in a window $1/a\gg 1/x\gg\Lambda_{QCD}$; we will 
call this definition as the X-space scheme. One can also
obtain $Z_\Gamma(\mu=1/x)$ directly in another continuum scheme,
such $\MSbar$,  by imposing,
\be\label{MSbar_ren}
{Z^{\MSbar}_\Gamma}^2(\mu=1/x) \, G_\Gamma(x)\,=\, 
G_\Gamma(\mu,x)^{(\MSbar)}
\ee
We have analyzed results for $O(a)$ improved Wilson fermions at
$\beta=6.45$ ($a^{-1}=3.87(19)$ GeV) on a $32^3\times
70$ lattice and overlap 
fermions at $\beta=6.00$ ($a^{-1}=2.00(10)$ GeV) on a $16^3\times
32$ volume. 
All the results presented here have been extrapolated  to the chiral limit 
(further information can be found in \cite{xspace_ours}). 

As an example of the general strategy adopted, we explain the analysis for the
vector-vector correlator with Wilson fermions. 
\begin{figure}[t!]
\bc
\begin{minipage}{3.6cm}
\includegraphics*[scale=0.61]{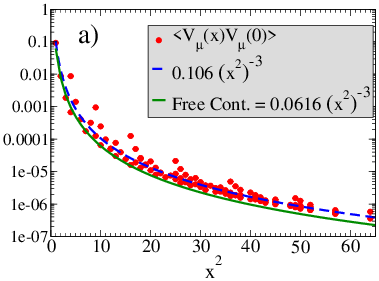}
\end{minipage}
\hfill
\begin{minipage}{3.6cm}
\includegraphics*[scale=0.63]{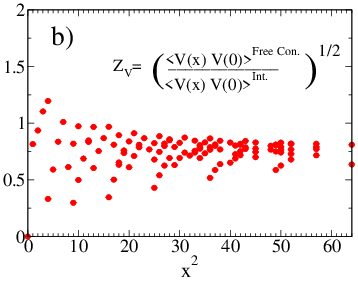}
\end{minipage}
\begin{minipage}{3.6cm}
\includegraphics*[scale=0.63]{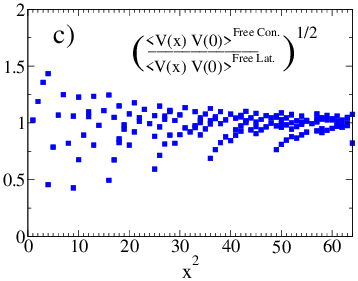}
\end{minipage}
\hfill
\begin{minipage}{3.6cm}
\includegraphics*[scale=0.63]{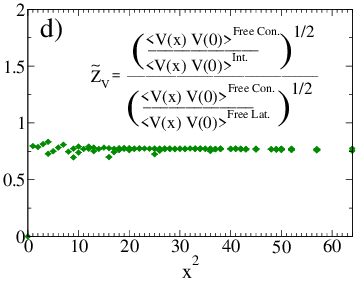}
\end{minipage}
\ec
\mbox{}\\[-1.9cm]
\caption{\label{fig_1}}
\mbox{}\\[-1.5cm]
\end{figure}
In Fig. \ref{fig_1}a) we show the dependence of the correlator $\langle
O_V(x) O_V^\dagger(0) \rangle$ on $x^2$. The na\"\i ve dimensional behaviour (${x^2}^{-3}$) is well satisfied 
but the data points are quite dispersed (see also Fig. \ref{fig_1}b)). 
To clarify the origin of these
effects we show in Fig. \ref{fig_1}c) the ratio between the
free correlator on the lattice  and in the continuum,
both in infinite volume (note that in this case the expected value is 1 up to $O(a)$ effects). Comparing both
figures we observe that the tree level calculation manifests the same pattern that the interacting case.
These considerations together with similar analysis (see \cite{xspace_ours}), brought us to conclude that 
we are observing discretization effects, which we propose to reduce considering the following 
ratio:
\be\label{Z_rat:def}
Z^2_\Gamma(\mu=1/x) \!=\! {G_\Gamma(x)^{(F.\: con.)}\over  G_\Gamma(x)^{(I.)}}\cdot
{G_\Gamma(x)^{(F.\: lat.)}\over  G_\Gamma(x)^{(F.\: con.)}}\:,
\ee
where, I., F. mean interacting and free respectively.
In this way we obtain the result plotted in Fig.
\ref{fig_1}d), which can be compared directly with Fig.
\ref{fig_1}b). It is clear that in this way a great part of $O(a)$ effects are
canceled out in the ratio of eq. \re{Z_rat:def}.
The same procedure has been carried out for the other bilinears and for the overlap data.

As commented above one can also obtain $Z_O$ directly in
another continuum scheme 
($\MSbar$), by imposing \re{MSbar_ren}.
Now one can evolve the RCs with the $\MSbar$ Renormalization Group (RG) factors,
up to a fixed
scale, say 2 GeV.
\be\label{MS:match}
Z^{\MSbar} (2\, \mathrm{GeV}) = {Z(x) \over W(\mu,2\, \mathrm{GeV}) R(\mu,1/x) }\:,
\ee
where $R(\mu, 1/x)$ is the matching between the X-space scheme and $\MSbar$, which
it has been calculated at NLO \cite{xspace_ours}.
 $W(\mu,2\, \mathrm{GeV})$ is the RGE factor.
In principle $\mu$ is arbitrary as far as $\mu\sim 1/x$ in order to avoid large logarithmic
contributions. This scale dependence 
should cancel in a RGI quantity up to truncation of the perturbative
series. In practice we have chosen  $\mu=1/x$ and $\mu=2/x$ and included the
difference as a systematic error due to NNLO effects.

\begin{figure}
\bc
\includegraphics*[scale=0.8]{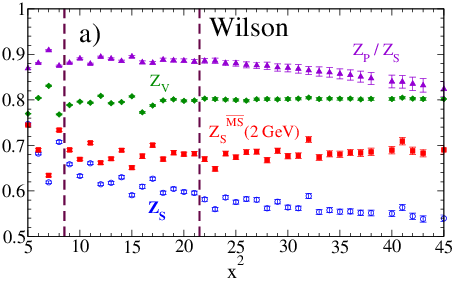}
\includegraphics*[scale=0.8]{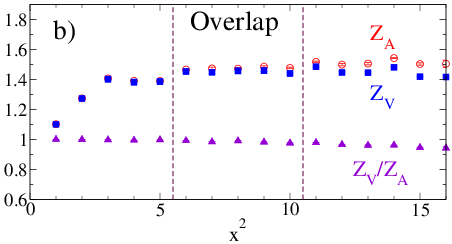}
\mbox{}\\[-1.4cm]
\caption{\label{fig_Zs}}
\ec
\mbox{}\\[-2.1cm]
\end{figure}
Our results are summarized in Fig. \ref{fig_Zs} for the
Wilson (\ref{fig_Zs}a) and overlap (\ref{fig_Zs}b) simulation, respectively. 
For the Wilson case we show, as an example, the
RCs for the scalar density and  vector current. We
also show the ratio $Z_P/Z_S$. This ratio (as well as $Z_V/Z_A$) allows to
extract information on the renormalization 
window, since it should be a plateau up to O(a) and NP effects. From Fig.
\ref{fig_Zs}a) we see that in the range $x^2=9\div 21$ we observe a reasonable
good plateau: For $x^2\le 9$ large $O(a)$ effects are present,
and above $x^2\sim 21$ the presence of NP effects begins to be visible (see 
$Z_P/Z_S$).
Therefore, we choose to obtain our final results for the RCs in
the above range. We also plot the scalar and vector RCs,
the first in $\MSbar$ at 2 GeV and in the X-space scheme.
For $Z_V$ we see a good plateau as one would expect. 
$Z_S(1/x)$ in the X-space  exhibits an anomalous dimension behaviour
as it should be. On the contrary in the value of $Z_S^{\MSbar}(2\,\mathrm{GeV})$ one expects that
the anomalous dimension cancels  (up to NNLO correction) and indeed a plateau is
observed. 
\begin{table*}[t!]
\caption{Wilson results and comparison with other methods: RI-MOM \cite{Becirevic:2002yv} and Schr\"odinger Functional
\cite{SF}.}
\label{table:1}
\newcommand{\m}{\hphantom{$-$}}
\newcommand{\cc}[1]{\multicolumn{1}{c}{#1}}
\renewcommand{\tabcolsep}{0.23pc} 
\renewcommand{\arraystretch}{1.2} 
\begin{tabular}{cllllll}
$Z_\Gamma$&\cc{$Z_V$}&\cc{$Z_A$}&\cc{$Z_P/Z_S$}&\cc{$Z_S$}&\cc{$Z_P$}&\cc{$Z_T$}\\
\hline
\hline
This work&$0.801(2)(18)(6)$&$0.833(2)(27)(6)$&$0.888(2)(8)$&
$0.702(4)(27)(23)$&$0.624(3)(19)(21)$&$0.895(2)(21)$\\
RI-MOM&$0.803(3)$&$0.833(3)$&$0.897(4)$&$0.679(8)$&$0.609(8)$&$0.898(6)$\\
SF&$0.808(1)$&$0.825(8)$&$0.912(9)$&$--$&$0.61(1)$&$--$\\
\hline
\end{tabular}
\end{table*}

For the overlap simulation, we show the currents RCs,
together with the ratio $Z_V/Z_A$. Since chirality holds exactly, one expect  
$Z_V=Z_A$ up to NP effects. From Fig. \ref{fig_Zs}b) one see that indeed this is the case for $x^2\lesssim 10$,
and we choose the range $6\div 10$ to obtain the values of $Z_V$, $Z_A$. 
Contrary to the Wilson results, no renormalization window has been found
for the other bilinears in the overlap case. This is due to the larger lattice
spacing (twice what we have for Wilson data) 
which consequently makes much narrower 
the renormalization window.
In Table 1 we present our results for the Wilson case and those with
other NP methods. 
All the RCs are in $\MSbar$ at $2$ GeV. The first error in our results is
statistic, the second one is an estimate of 
discretization errors and the last one comes from NNLO effects due to the
choice of the
scale (see above). 
For the overlap case we quote the results obtained for $Z_V$ and $Z_A$:
\be
Z_V=1.471(2)(10)(20)\:\, Z_A=1.494(2)(9)(20),
\ee
the tiny difference between the results for $Z_V$ and $Z_A$ is due to small NP
effects. In any case they are compatible within the errors. Our results
for $Z_A$ is comparing with 
$Z_A=1.55(4)$ obtained with the WI method in \cite{Giusti:2001yw}.
From Table 1 and the above results in the overlap case we conclude that although
the errors are, in most cases, larger than those 
obtained from other NP techniques, good agreement is found for all the cases which we can compare.
We stress that one could reduce the errors carring out a recursive volume
matching technique.
\begin{figure}
\bc
\mbox{}\\[-0.3cm]
\includegraphics*[scale=0.6]{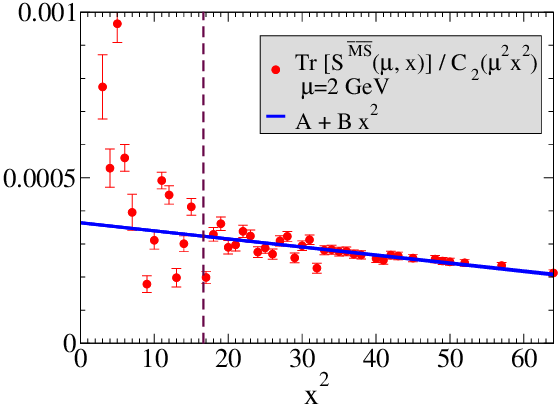}
\mbox{}\\[-1.3cm]
\caption{\label{fig_cond}}\mbox{}\\[-1cm]
\ec
\mbox{}\\[-2.3cm]
\end{figure}

We present a preliminary result for the chiral quark condensate, obtained in a
novel way. The idea is very simple and consists on studying the chiral limit of
the trace of the quark propagator in X-space together with its OPE:
\mbox{}\\[-.4cm]
\be\label{OPE_eq_1}
\mathrm{Tr}S(x)\sim C_1(x^2)\, {m\over x^2} -
C_2(x^2)\langle\bar \psi \psi\rangle \, +\, \cdots
\ee
\mbox{}\\[-.4cm]
We stress that in \re{OPE_eq_1}, once the l.h.s. is renormalized, the r.h.s.
gets renormalized too, and no other divergence (apart from the natural one at
$x=0$) is present, thus only the renormalization constant for the
quark field is needed. 
We have computed the needed Wilson coefficients factors in eq. \re{OPE_eq_1}.
From eq. \re{OPE_eq_1} and in the chiral limit one can obtain an estimate of the
chiral condensate. This is 
shown in Fig. \ref{fig_cond}. From that figure we see large discretization
effects for $x^2\le 15$. One can see also that there is a contribution from the
next order term which goes as $x^2$. From a linear fit we obtain our
(preliminary) result:
\be
\langle \bar \psi \psi \rangle^{\MSbar}(2\,\mathrm{GeV})\,=\,-
(276(5)(13)\,\mathrm{MeV})^3\:,
\ee
in good agreement with other lattice determinations. The first error is
statistic and the second is due to the uncertainty in the value of $a^{-1}$.
\mbox{}\\[-0.56cm]

\end{document}